\documentclass[%
 reprint,
preprintnumbers,
longbibliography,
 amsmath,amssymb,
 aps,
 prc,
 superscriptaddress,
]{revtex4-2}

\usepackage{xcolor}
\usepackage{graphicx} 
\usepackage{subfigure}
\usepackage{dcolumn} 
\usepackage{bm}
\usepackage{float}
\usepackage[columnwise,running]{lineno}
\usepackage{hyperref}
\usepackage{color}
\definecolor{orange}{cmyk}{0.,0.353,1.,0.}    % orange

\usepackage[outdir=./]{epstopdf}
\begin{document}

\title{Fourier coefficients of noninterdependent collective motions in heavy-ion collisions}
\author{Zhiwan Xu}\email{zhiwanxu@physics.ucla.edu}
\affiliation{Department of Physics and Astronomy, University of
  California, Los Angeles, California 90095, USA}
  \author{Gang Wang}\email{gwang@physics.ucla.edu}
\affiliation{Department of Physics and Astronomy, University of
  California, Los Angeles, California 90095, USA} 
\author{Aihong Tang} \affiliation{Brookhaven National Laboratory, Upton, New York 11973} 
  \author{Huan Zhong Huang} \affiliation{Department of Physics and
  Astronomy, University of California, Los Angeles, California 90095,
  USA} \affiliation{Key Laboratory of Nuclear Physics and Ion-beam
  Application (MOE), and Institute of Modern Physics, Fudan
  University, Shanghai-200433, People’s Republic of China}

%\date{} %leave blank

\begin{abstract}
We present a scenario in heavy-ion collisions where different modes of collective motions are noninterdependent, driven by factorized actions in the created nuclear medium. Such physics mechanisms could each dominate at a distinct evolution stage, or coexist simultaneously. If the probability of particle emission is modulated by each
nondependent collective motion with a single-harmonic Fourier expansion, the particle azimuthal distribution
should be the product of all these expansions. Consequently, nonleading cross terms between collectivity modes
appear, and their contributions to experimental observables could be significant. In particular, we argue that
the chiral magnetic effect (CME) and elliptic flow can develop separately, with their convolution affecting the
observable that is sensitive to the shear-induced CME. We will use the event-by-event anomalous-viscous fluid
dynamics model to illustrate the effects of this scenario. Besides giving insights into searches for the CME, we
also propose feasible experimental tests based on conventional flow harmonics, and demonstrate the emergence
of nonleading cross terms with a multiphase transport model.

%Consequently, cross terms between different harmonics will appear, and the  experimental observables based on a linear Fourier series will deviate from the true coefficients in the factorized form.

\begin{description}
\item[keywords]
chiral magnetic effect; elliptic flow; heavy-ion collision; non-interdependent
\end{description}
\end{abstract}

\maketitle
%\section{Introduction}
%\label{sec:intro}

\begin{figure}[tb]
\includegraphics[width=0.4\textwidth]{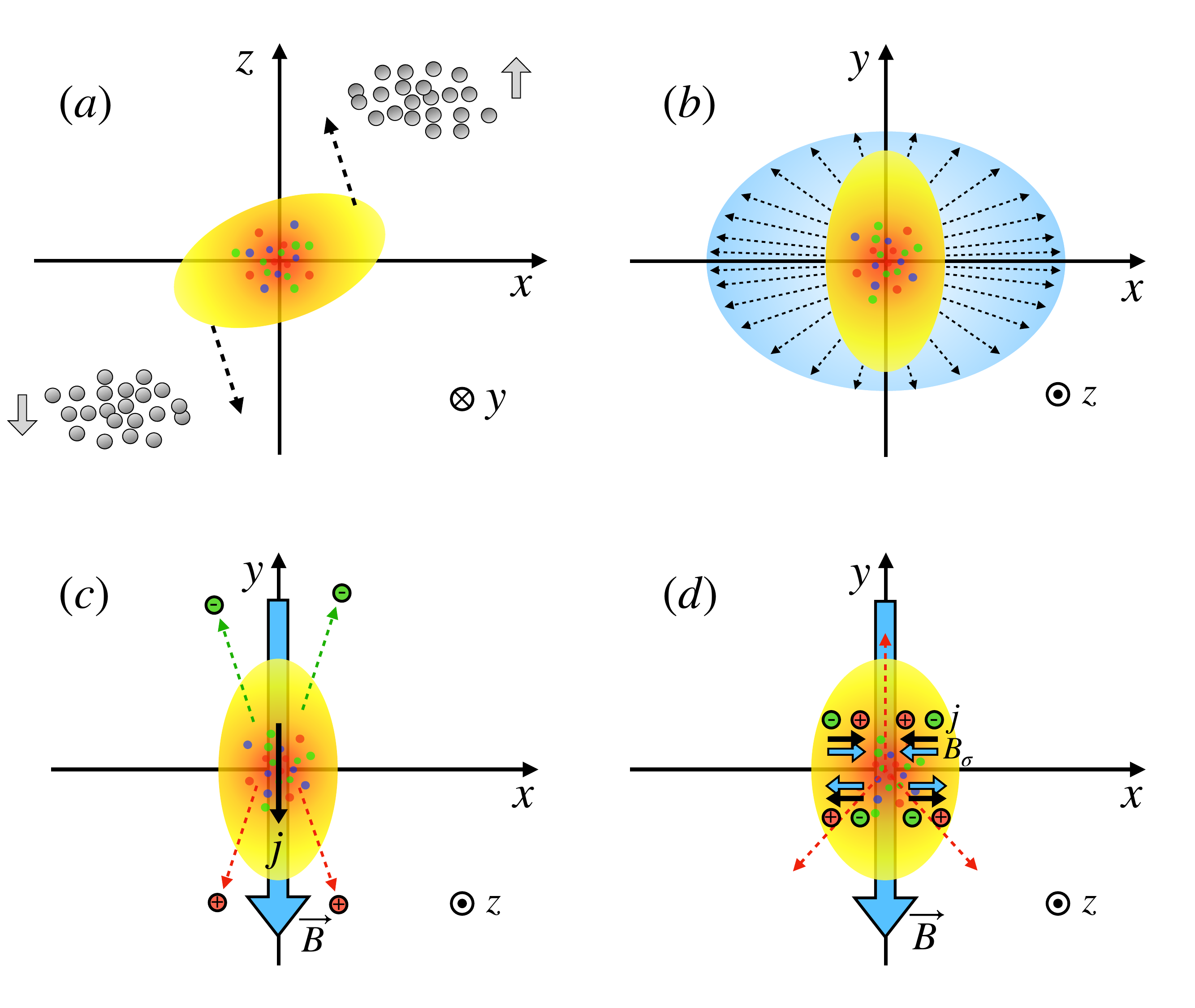}
\caption{Illustration of collective motions in heavy-ion collisions: (a) directed flow, (b) elliptic flow, (c) the CME-induced electric current $j$ along the $\vec{B}$ field, and (d) the siCME-induced charge-dependent  triangular emission. The dashed arrows represent  produced-particle momenta. $B_\sigma$ denotes the new component of the $\vec{B}$ field caused by its interplay with shear flow.}
\label{fig:setup}
\end{figure}

In high-energy heavy-ion collisions, the emission pattern
of final-state particles reveals different collectivity modes of
the created nuclear medium. Figure~\ref{fig:setup} illustrates a few examples of collective motions at midrapidities in noncentral
collisions. A particular concept is the reaction plane, spanned
by impact parameter ($x$ axis) and beam momenta ($z$ axis).
(a) In the reaction plane, a slightly tilted participant region
~\cite{tilt} violates the boost invariance, and leads to a rapidity-odd
emission of produced particles, known as directed flow ($v_1$).
(b) When viewed along the beam line, the almond-shaped
overlap zone in coordinate space is transformed via a hydrodynamic expansion~\cite{hydro} into a rapidity-even nondegeneracy
between in-plane and out-of-plane emissions, called elliptic
flow ($v_2$). (c) The chiral magnetic effect (CME)~\cite{CME} induces
an out-of-plane electric charge separation ($a_1^\pm$), provided that
a quark chirality imbalance emerges from the chiral anomaly
~\cite{Anomaly}, and an intense magnetic field ($\vec{B}$) is generated by protons
from the colliding nuclei~\cite{Bfield}. (d) Recently, a higher-order effect, the shear-induced CME (siCME)~\cite{siCME} has been proposed,
in which the combination of magnetic field and hydrodynamic
shear creates a charge-dependent triangular flow ($a_3^\pm$).

To quantify the collective motions in a given kinematic region, it is convenient to decompose
the azimuthal angular ($\varphi$) distribution of produced particles in each collision with a Fourier series~\cite{flow}:
\begin{equation}
\frac{2\pi}{N^\pm}\frac{dN^\pm}{d\varphi} = 1+ \sum_{n=1}^{\infty}2a_n^\pm\sin n\Delta \varphi+ \sum_{n=1}^{\infty} 2v_n^\pm\cos n\Delta \varphi ,
\label{Eq: Fourier expansion}
\end{equation}
where $\Delta \varphi$ is the azimuthal angle of a particle relative to the reaction plane, and the superscripts $+$, $-$ indicate the charge signs. For simplicity, we will omit these superscripts in the following discussions, and  express them only where necessary.
The coefficients $a_n\equiv\langle \sin n\Delta\varphi\rangle$ and $v_n\equiv\langle \cos n\Delta\varphi\rangle$  are experimentally obtainable by averaging over particles of interest and over events. While Fourier-expanding a probability distribution is always a useful approach, establishing the relationship between the Fourier coefficients and underlying physics mechanisms can be nontrivial. Storing coefficients in an orthonormal basis implies that each collectivity mode is treated as a nondependent action on the final-state particle distribution.
However, this leads to a logical inconsistency. If the collectivity modes are orthogonal or noninterdependent,   each action should modulate the probability of particle emission with its own single-harmonic  ($\tilde{a}_n$
or $\tilde{v}_n$) Fourier expansion. According to the rule of multiplication for independent actions, it is the product, rather than the sum, of these short expansions that describes the particle azimuthal distribution. 
Then, the $a_n$ ($v_n$) measured from a long linear Fourier expansion as in Eq.~(\ref{Eq: Fourier expansion}) may not be independent of each other, and may not fully match the true $\tilde{a}_n$ ($\tilde{v}_n$) of the pertinent physics process. 

In general, factorized actions, such as distinct physics mechanisms, could occur concurrently, but do not rely on each other's existence during evolution. 
As long as the collective motions have separate evolution paths to affect particle emission, we consider them to be not coordinated with each other and  approximate the particle distribution with the factorized form of Fourier expansions. For example, elliptic flow can be developed regardless of the presence of the CME, and vice versa. 
Then, $\Tilde{a}_1$ and $\Tilde{v}_2$ should appear in two separate Fourier expansions, $(1+2\Tilde{a}_1 \sin\Delta\varphi)$ and $(1+2\Tilde{v}_2 \cos2\Delta\varphi)$, respectively, and their product specifies the final-state particle distribution.
Compared with Eq.~(\ref{Eq: Fourier expansion}), the factorized form provides a nonleading cross term:
\begin{eqnarray}
& &4\Tilde{a}_1 \Tilde{v}_2\sin\Delta\varphi\cos2\Delta\varphi \nonumber \\
&=& -2\Tilde{a}_1 \Tilde{v}_2\sin\Delta\varphi
+2 \Tilde{a}_1\Tilde{v}_2\sin3\Delta\varphi.
\label{Eq:2}
\end{eqnarray}
In this case, the $a_1$ and $a_3$  manifested in the linear Fourier expansion deviate from the true $\Tilde{a}_1$ and $\Tilde{a}_3$, respectively.
 
Under the hypothesis that the actions responsible for collective motions can be factorized, the particle azimuthal distribution would be expressed naturally as
\begin{equation}
\frac{2\pi}{N^\pm}\frac{dN^\pm}{d\varphi} = \prod_{n=1}^{\infty}(1+2\tilde{a}_n^\pm\sin n\Delta \varphi) \prod_{n=1}^{\infty} (1+ 2\tilde{v}_n^\pm\cos n\Delta \varphi).
\label{Eq: Fourier expansion2}
\end{equation}
Note that Eq.~(\ref{Eq: Fourier expansion2}) is not a replacement of 
Eq.~(\ref{Eq: Fourier expansion}), but they are two representations of the same distribution with different emphases, and both are  applicable to real-data analyses. The coefficients in the former expansion represent the  strengths of the collective motions driven by factorized actions, whereas those in the latter denote the final emergent collectivity modes. Typically, flow and CME measurements are conducted using multiparticle correlations based on the definition in Eq.~(\ref{Eq: Fourier expansion}), and the results thus obtained manifest $v_n$ ($a_n$), which can be readily demodulated into $\tilde{v}_n$ ($\tilde{a}_n$).

In reality, the difference between $a_n$ and $\tilde{a}_n$
or between $v_n$ and $\tilde{v}_n$
is negligible for many harmonics. For example, the magnitude of $\tilde{a}_n\tilde{a}_m$ could be much smaller  than $\tilde{v}_{n+m}$ or $\tilde{v}_{|n-m|}$. We will focus on three coefficients: $\tilde{a}_1$, $\tilde{a}_3$ and $\tilde{v}_2$, and study how they are related to their counterparts, $a_1$, $a_3$ and $v_2$. Directed flow is not included, because $v_1$ is a rapidity-odd function in symmetric collisions, and its rapidity-integrated contribution to other coefficients will be zero in most cases. Now,  Eq.~(\ref{Eq: Fourier expansion2}) takes a specific form:
\begin{eqnarray}
\frac{2\pi}{N^\pm}\frac{dN^\pm}{d\varphi} &\propto& (1+2\tilde{a}_1^\pm \sin\Delta\varphi)  \times(1+2\tilde{a}_3^\pm \sin3\Delta\varphi)  \nonumber \\
& & \times(1+2\tilde{v}_2^\pm \cos2\Delta\varphi).
\label{Eq: Fourier expansion3}
\end{eqnarray}
By comparing Eqs.~(\ref{Eq: Fourier expansion}) and (\ref{Eq: Fourier expansion3}), we find the following connections between phenomena and  noumena,
\begin{eqnarray}
a_1 &=& \tilde{a}_1 - \tilde{a}_1\tilde{v}_2  +  \tilde{a}_3\tilde{v}_2 , \label{Eq:a1}\\
a_3 &=& \tilde{a}_3 + \tilde{a}_1 \tilde{v}_2, \label{Eq:a3}\\
v_2 &=& \tilde{v}_2 + \tilde{a}_1 \tilde{a}_3.
\end{eqnarray}
Here we ignore any higher-order term involving $\tilde{a}_1 \tilde{a}_3\tilde{v}_2$.

Since the magnitude of $\tilde{a}_1 \tilde{a}_3$ is typically lower than that of $v_2$ by a few orders of magnitude, $v_2$ and $\tilde{v}_2$ are almost the same. We will abandon $\tilde{v}_2$, and only use $v_2$ in the following discussions.
Given that the siCME-induced $\tilde{a}_3$ is much smaller than the CME-induced $\tilde{a}_1$,  Eq.~(\ref{Eq:a1})  indicates that the observed $a_1$ roughly equals $\tilde{a}_1(1-v_2)$. 
Furthermore, Eq.~(\ref{Eq:a3}) asserts that the observed $a_3$ contains a contribution of $\tilde{a}_1 v_2$ on top of the primordial $\tilde{a}_3$, if any. 
For completeness, we  also express $\tilde{a}_1$ and $\tilde{a}_3$
in terms of experimental observables:
\begin{eqnarray}
\tilde{a}_1 &=& \frac{a_1-a_3v_2}{1-v_2-v_2^2}, \\
\tilde{a}_3 &=& \frac{a_3-a_1v_2-a_3v_2}{1-v_2-v_2^2}.
\end{eqnarray}

%\section{EBE-AVFD model}

We use the Event-by-Event Anomalous-Viscous Fluid Dynamics (EBE-AVFD) model~\cite{AVFD1,AVFD2,AVFD4} to test our inferences from  factorized actions via relations derived in Eqs.~(\ref{Eq:a1}) and (\ref{Eq:a3}).
The EBE-AVFD event generator simulates the dynamical CME
transport for $u$, $d$ and $s$ quarks in addition to  the hydrodynamically expanding viscous medium in heavy-ion collisions, and properly handles local charge conservation and resonance decays. We have analyzed $5.8\times10^7$ events of Au+Au collisions at $\sqrt{s_{NN}} = 200$ GeV in the 30--40\% centrality range, using the same settings and input parameters as adopted in Ref.~\cite{AVFD5}. For simplicity, we will use the true reaction plane to perform the simulation analysis, and ignore the possible fluctuation effects concerning the observables of elliptic flow and the CME.

The initial conditions for entropy density ($s$) profiles and for electromagnetic field vary in accordance with the event-by-event nucleon configuration from the Monte Carlo Glauber simulations~\cite{glauber}. The  chirality charge density ($n_5$) is implemented in the form of $n_5/s$, which controls the strength of the CME transport. In this study, we take a modest value of  $n_5/s = 0.1$, the same as used in Ref.~\cite{siCME}. 

The medium expansion is managed by the VISH2+1 simulation package~\cite{AVFD3}, 
which is a boost-invariant hydrodynamics framework.
Consequently, directed flow vanishes, and elliptic flow is a major collectivity mode.
In these EBE-AVFD calculations, magnetic field is set to last long enough, so that for a substantial time period the CME does coexist with the hydrodynamic formation of elliptic flow. However, the dynamical CME transport is governed by anomalous hydrodynamic equations as a linear perturbation on top of the medium flow background, since the back-reaction of finite chiral quark densities is negligible in collisions at $\sqrt{s_{NN}} = 200$ GeV~\cite{AVFD1}. Therefore, the two modes of collective motions featuring $\tilde{a}_1^\pm$ and $v_2$, respectively, develop independently of each other, and satisfy the generalized criterion for noninterdependent collective motions.
We do not invoke the siCME in these simulations, and thus $\tilde{a}_3^\pm$ is zero. Accordingly, Eqs.~(\ref{Eq:a1}) and (\ref{Eq:a3}) become as simple as
\begin{eqnarray}
a_1 &=& \tilde{a}_1 (1- v_2) , \label{Eq:a1_new}\\
a_3 &=& \tilde{a}_1 v_2. \label{Eq:a3_new}
\end{eqnarray}

%\section{Results and Discussions}

\begin{figure}[tb]
\includegraphics[width=0.48\textwidth]{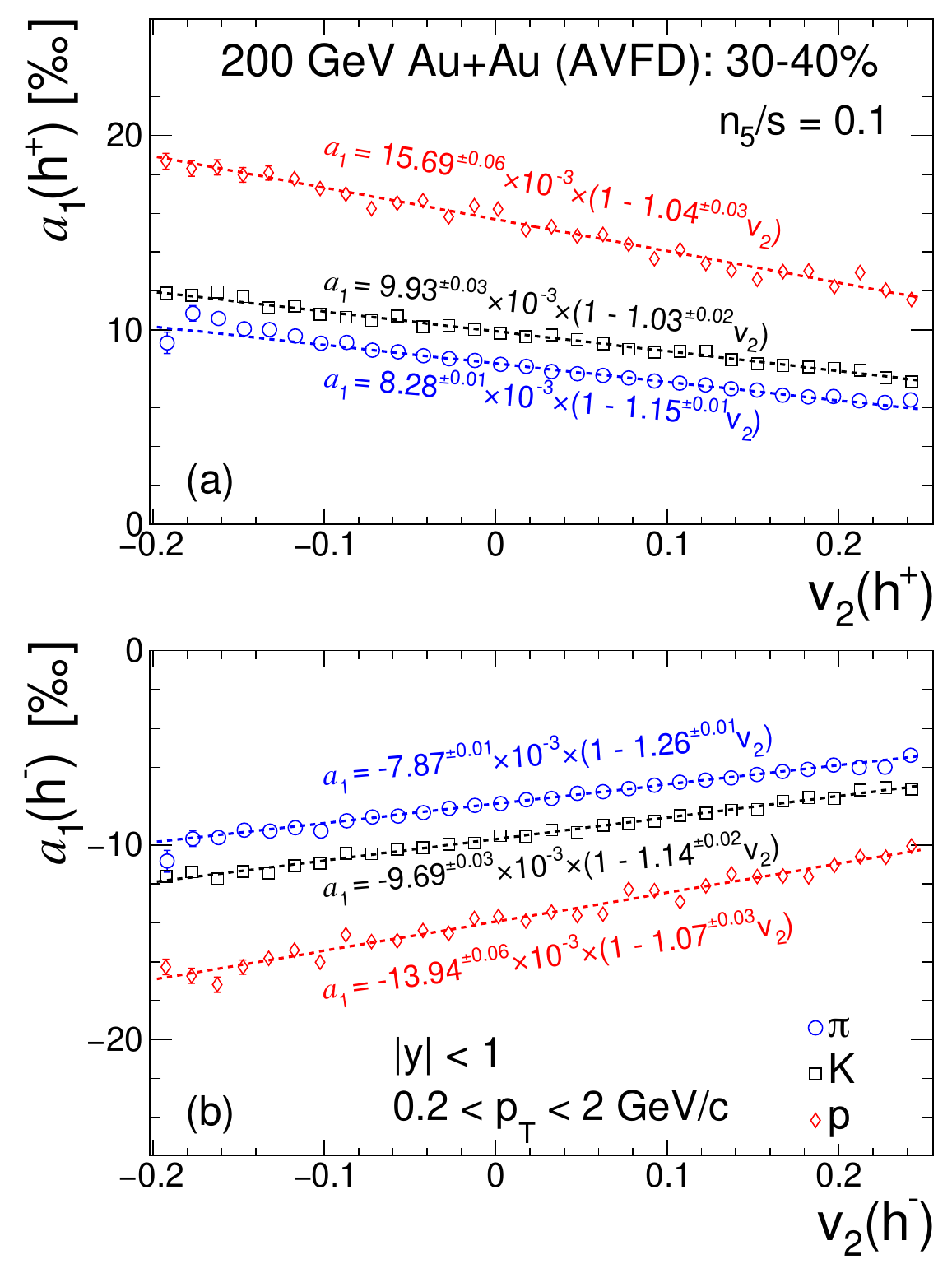}
\caption{Correlations between $a_1$ and $v_2$ calculated on an
event-by-event basis from EBE-AVFD simulations of 30--40\% Au+Au collisions at $\sqrt{s_{NN}} = 200$ GeV for (a) $\pi^{+}$, $K^+$ and $p$, and for (b) $\pi^{-}$, $K^-$ and $\bar{p}$. The fit function, $a_1(v_2) = a_1(0)\times(1 + C v_2)$, returns  $C$ close to $-1$ for all the cases.}
\label{fig:a1}
\end{figure}

Figure~\ref{fig:a1}
shows the observed $a_1$ vs $v_2$ from EBE-AVFD simulations of  Au+Au collisions at $\sqrt{s_{NN}} = 200$ GeV in the centrality interval of 30--40\% for (a) $\pi^{+}$, $K^+$ and $p$, and for (b) $\pi^{-}$, $K^-$ and $\bar{p}$. Both $a_1$ and $v_2$ are calculated on an
event-by-event basis  within the rapidity range of $|y|<1$ and the transverse momentum range of $0.2 < p_T < 2$ GeV/$c$. For all the particle species, 
a linear correlation is present between $a_1$ and $v_2$. The fit function (dashed lines), $a_1(v_2) = a_1(0)\times(1 + C v_2)$, renders the  parameter $C$ close to $-1$ for all the cases, seemingly supportive of Eq.~(\ref{Eq:a1_new}). However, we cannot claim unambiguous evidence for the scenario of factorized actions based solely on this slope parameter, due to the trigonometric identity 
\begin{equation}
\sin^2 \Delta\varphi \equiv (1 -\cos2\Delta\varphi)/2.    \label{eq:triag}
\end{equation}
Even when both $a_1$ and $v_2$ are averaged over particles in each event before being correlated, the correlation could still be dominated by the mathematical relation. In addition, there could be an anticorrelation between the $\tilde{a}_1$ magnitude itself and $v_2$, especially for pions. This higher-order effect may arise from resonance decays, since secondary particles will smear and dilute the $\tilde{a}_1$ value averaged over all pions~\cite{AVFD1}, while inheriting a larger $v_2$ from resonances at higher $p_T$~\cite{decay1, decay2}.
The $a_1(0)$ or $\tilde{a}_1$ values retrieved from the fits also exhibit a particle-species dependence, which could be partially explained by the mean $p_T$ effect. Similar to $v_2$, the $\tilde{a}_1$ magnitude  increases with $p_T$ at the low-$p_T$ region~\cite{AVFD1}, and pions, kaons, and protons form an ascending order of mean $p_T$. The quark coalescence mechanism~\cite{coalescence} could also play a role by making the $\tilde{a}_1$ magnitude larger for baryons than for mesons at the intermediate-$p_T$ region.

\begin{figure}[tbhp]
\includegraphics[width=0.48\textwidth]{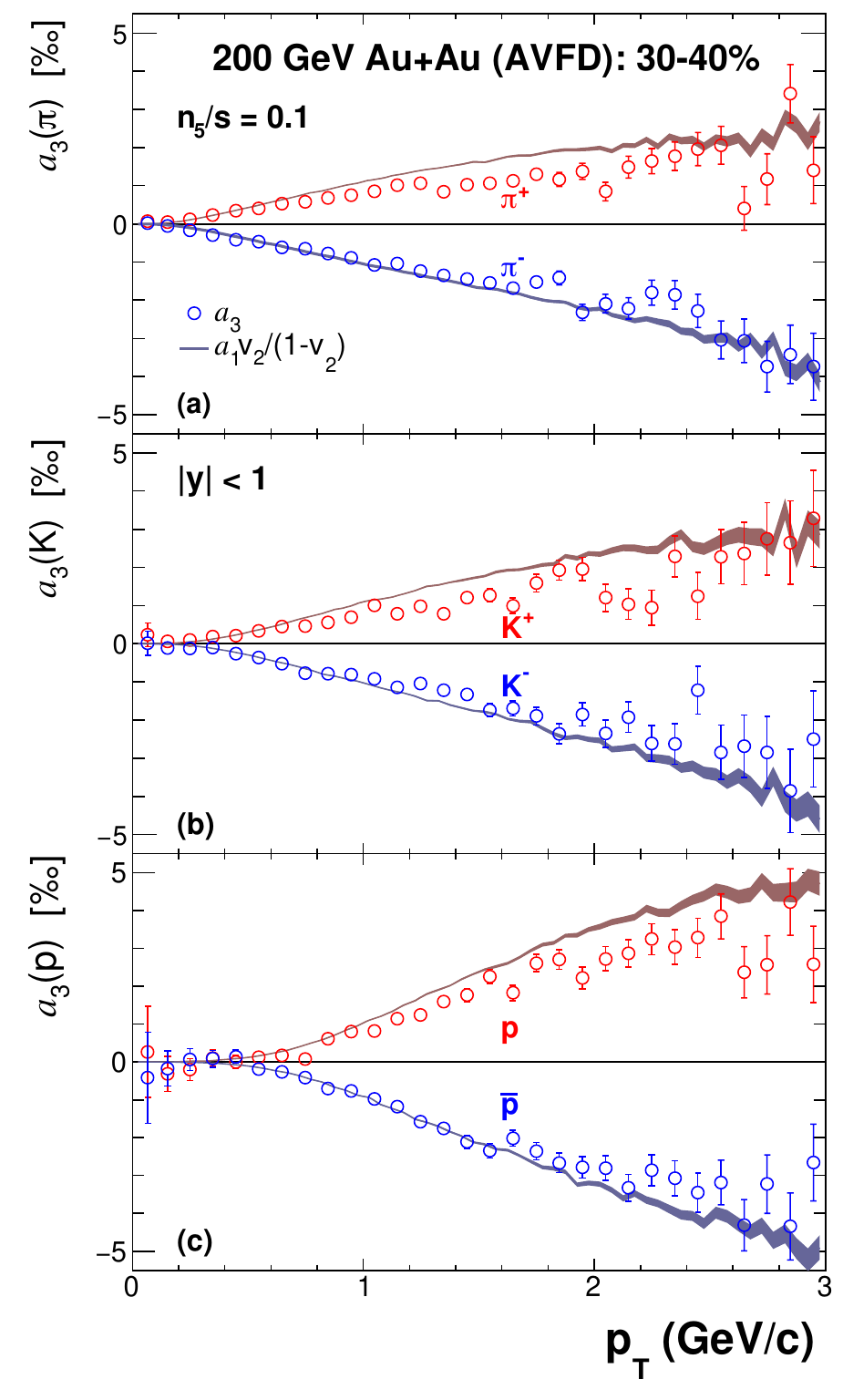}
\caption{EBE-AVFD simulations of $a_3$ as a function of $p_T$ in 30--40\% Au+Au collisions at 200 GeV for (a) $\pi^{+}$ and $\pi^{-}$, for (b) $K^+$ and $K^-$, and for (c) $p$ and $\bar{p}$. In comparison, the values of $a_1v_2/(1-v_2)$ are shown with shaded bands.}
\label{fig:fig2a3}
\end{figure}

Figure~\ref{fig:fig2a3} shows EBE-AVFD calculations of the observed $a_3$ as a function of $p_T$ in 30--40\% Au+Au collisions at 200 GeV for (a) $\pi^{+}$ and $\pi^{-}$, for (b) $K^+$ and $K^-$, and for (c) $p$ and $\bar{p}$.
Although the $a_3$ magnitudes for pions and protons are comparable to the corresponding predictions for the siCME-induced $a_3$ (as shown in the upper panel of Fig.~3 in Ref.~\cite{siCME}),  our simulations do not entail the siCME. 
On the contrary, we depict $a_1v_2/(1-v_2)$ or $\tilde{a}_1v_2$ with shaded bands, which  describe the trend and the magnitude of $a_3$ reasonably well for all the particle species under study.
Therefore, the simulations support Eq.~(\ref{Eq:a3_new}), and verify another imprint of factorized actions. At $p_T > 1$ GeV/$c$, $a_3$ seems to have a smaller magnitude than $\tilde{a}_1v_2$, which could be partially explained by the aforementioned  anti-correlation between $\tilde{a}_1$ and $v_2$. The gradual breakdown of hydrodynamics  towards higher $p_T$ could also add to this discrepancy, since collective motions start to collapse.

\begin{figure}[tbhp]
\includegraphics[width=0.48\textwidth]{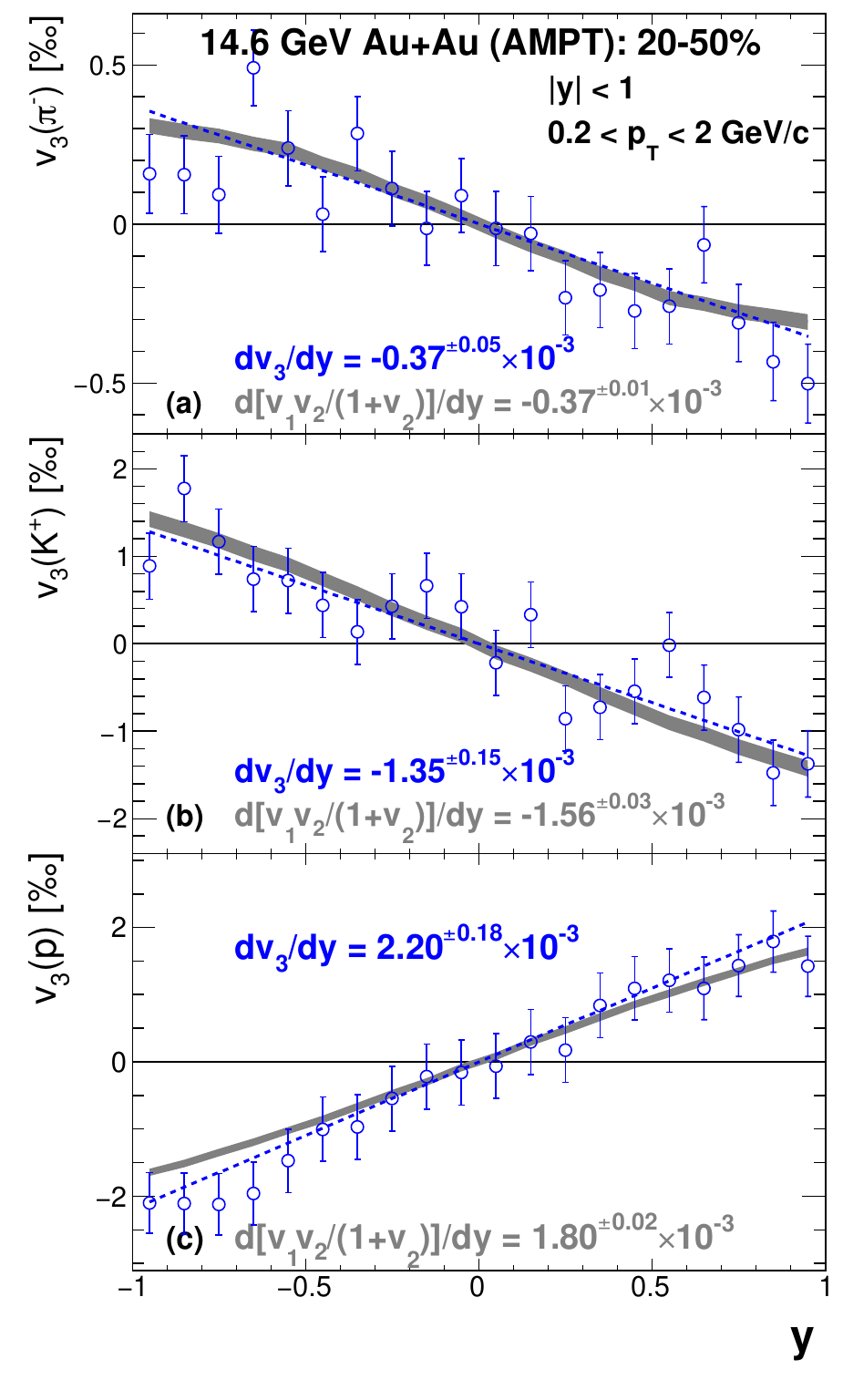}
\caption{AMPT simulations of  $v_3$ vs $y$  for (a) $\pi^-$, (b) $K^+$, and (c) $p$ in 20--50\% Au+Au collisions at 14.6 GeV. For comparison, the values of $v_1v_2/(1+v_2)$ are shown with shaded bands. Linear fits are used to extract the slopes.
}
\label{fig:fig3}
\end{figure}

While the search for the CME and the siCME continues as a subject of intensive investigation, we propose a feasible approach to assess the factorized-form approximation   involving only the $v_n$ coefficients.
If directed flow and elliptic flow are approximately noninterdependent in nature, the product of 
$(1+2\Tilde{v}_1 \cos\Delta\varphi)$ and $(1+2\Tilde{v}_2 \cos2\Delta\varphi)$ yields a non-leading cross term, $4\tilde{v}_1 \tilde{v}_2 \cos\Delta\phi\cos2\Delta\phi = 2\tilde{v}_1 \tilde{v}_2(\cos\Delta\phi+\cos3\Delta\phi)$, similar to Eq.~(\ref{Eq:2}). Therefore, the  $v_1$ observed with a linear Fourier expansion will be the true $\tilde{v}_1$ scaled by a factor of $(1+\tilde{v}_2)$, and the observed $v_3$ will contain a rapidity-odd component of 
$\tilde{v}_1 \tilde{v}_2$ on top of the existing rapidity-even $\tilde{v}_3$, if any.
With respect to the reaction plane, $\tilde{v}_3$ is likely to be zero because  the triangular anisotropy in the initial collision geometry is dominated by event-by-event fluctuations, and essentially decouples from the reaction plane~\cite{v3}. Thus, the observation of a  $v_3^{\rm odd}$ component establishes a signature of the factorized Fourier expansions.
Along the same line of argument,
the $v_2$ observed with the linear Fourier expansion will be
the true $\tilde{v}_2$ scaled by $(1+\tilde{v}_4)$.

To test the aforementioned ideas, we exploit a multi-phase transport (AMPT) model~\cite{ampt1,ampt2}, whose string melting version~\cite{ampt2,ampt3}  reproduces particle spectra and flow reasonably well at both RHIC and LHC energies~\cite{ampt4}. We  simulate Au+Au collisions at $\sqrt{s_{NN}}= 14.6$ GeV/$c$, where directed flow is prominent. 
Since the correlations between $v_1$ and $v_2$ and between $v_2$ and $v_4$ could be affected by pure mathematical relations similar to Eq.~(\ref{eq:triag}), we focus on $v_3^{\rm odd}$.
Figure~\ref{fig:fig3} presents AMPT calculations of  $v_3$ vs $y$  for (a) $\pi^-$, (b) $K^+$, and (c) $p$ in the centrality range of 20--50\%. We refrain from combining the results for particles and their antiparticles due to sizeable differences arising from transported-quark effects at this beam energy~\cite{transport}.  
In each panel, we also include a shaded band to depict $v_1v_2/(1+v_2)$ for comparison. For all studied particle species, the slope, $dv_3/dy$, is statistically significant and consistent with that of $v_1v_2/(1+v_2)$, supporting the  scenario of noninterdependent collective motions.

In summary, we stipulate that collective motions in high-energy heavy-ion collisions may be noninterdependent, and that the particle azimuthal distribution can take a factorized form that complements the widely used long linear Fourier series. This scheme is more self-consistent, better captures the genuine strength of each collectivity mode, and makes new predictions based on nonleading cross terms. In the experimental extraction of Fourier coefficients, concerns of factorization or rather lack thereof have been raised from the viewpoint of nonflow~\cite{Ollitrault2013,Lim2019,Kikola2012} or decorrelation~\cite{Bozek2018,Barbosa2021}, but none of them involves noninterdependent collective motions or their effects on the particle distribution.
As a concrete example, we focus on the extra cross terms between the CME-induced $\tilde{a}_1$, the siCME-induced $\tilde{a}_3$, and elliptic flow $v_2$, and make two predictions. First,  the presence of a finite elliptic flow will scale $\tilde{a}_1$ by a factor of $(1-v_2)$, and this effect is more important at higher $p_T$, where $v_2$ is larger. 
Since most CME-sensitive observables contain  $a_1^2$,  the corresponding reduction factor should be about $(1-v_2)^2$.
Second, as confirmed by the EBE-AVFD calculations, the observed $a_3$ receives a sizeable contribution from $\tilde{a}_1 v_2$, which complicates the interpretation of this siCME sensitive observable.
Nevertheless, a finite $a_3$, if confirmed, constituents a strong evidence of the CME, whether it originates from $\tilde{a}_3$ or $\tilde{a}_1 v_2$ or both.
We have also proposed a test to examine noninterdependent collective motions using the $v_n$ coefficients, and the AMPT  simulations corroborate the prediction of the rapidity-odd component of $v_3$. The universality of the assumption regarding factorized actions can be investigated through analyzing real data collected from RHIC and the LHC, which will enhance our understanding of the collective motions.

\begin{acknowledgments}
{The authors thank Shuzhe Shi and Jinfeng Liao for providing the EBE-AVFD code and for many fruitful discussions on the CME.
We also thank Yufu Lin for generating the EBE-AVFD events. We thank Zi-Wei Lin and Guo-Liang Ma for providing the AMPT code.
We are especially grateful to Zhongling Ji and Yicheng Feng
for the fruitful discussions.
Z.X., G.W., and H.Z.H. are supported by the U.S. Department 277 of Energy under Grant No. DE-FG02-88ER40424 and by the 278 National Natural Science Foundation of China under Contract 279 No. 1835002. A.H.T. is supported by the U.S. Department of 280 Energy under Grants No. DE-AC02-98CH10886 and No. DE- 281 FG02-89ER40531
}
\end{acknowledgments}

{}

\end{document}